\documentclass[conference]{IEEEtran}
\usepackage{pifont}
\usepackage{times,amsmath,color,amssymb,graphicx,epsfig,cite,psfrag,subfigure,algorithm,balance}
\usepackage{amsfonts,pifont,enumerate,cases}
\usepackage{mathrsfs} % \mathcal 宏包支持
\usepackage[table]{xcolor} % 用于表格各行颜色
\usepackage{verbatim} % 与\begin{comment}...\end{comment}配合用户注释
\usepackage{lettrine} % \lettrine[lines=2]{L}{arge-scale}
\usepackage{bm}
\usepackage{geometry}
\newtheorem{property}{Property}

\newtheorem{remark}{\underline{Remark}}

\newtheorem{proof}{Proof}

\newcommand{\tu}{\textup}
\IEEEoverridecommandlockouts   % Conference 里面使\thanks 命令生效的语句
\begin{document}

%\newgeometry{top=2cm,bottom=4.29cm,left=1.4cm,right=1.4cm}

\title{Spatial-Temporal BEM and Channel Estimation Strategy for Massive MIMO Time-Varying Systems}
\author{Hongxiang Xie$^*$, Feifei Gao$^*$, Shun Zhang$^\dag$,  and Shi Jin$^\ddag$\\
$^*$ Tsinghua National Laboratory for Information Science and Technology (TNList), Beijing\\
$^\dag$ State Key Laboratory of Integrated Services Networks, Xidian University\\
$^\ddag$ National Communications Research Laboratory, Southeast University
\thanks{This work was supported in part by the National Natural Science Foundation of China under Grant \{61422109, 61531011\}.
}
}
\maketitle
\thispagestyle{empty}
\begin{abstract}
This paper proposes a new channel estimation scheme for the multiuser
massive multiple-input multiple-output (MIMO) systems in time-varying environment.
We introduce a discrete Fourier transform (DFT) aided spatial-temporal
basis expansion model (ST-BEM) to reduce the effective dimensions of uplink/downlink channels, such that
training overhead and feedback cost could be greatly decreased. The newly proposed ST-BEM
is suitable for both time division duplex (TDD) systems and
frequency division duplex (FDD) systems thanks to the angle reciprocity,
and can be efficiently deployed by fast Fourier transform (FFT).
Various numerical results have corroborated the proposed studies.
\end{abstract}

\enlargethispage{-0.3cm}
\section{Introduction}

Channel estimation has been a major challenge for massive multiple-input multiple-output (MIMO) system, where most existing works,
e.g. \cite{marzetta2010noncooperative}, focus on time-invariant environments. However, in many mobile environment, the
time varying channel estimation should also be considered so as to improve the accuracy of data detection.

Since channel parameters cannot change in a sudden way in time domain, one may
expect correlation among the time varying channel parameters. Exploiting this fact,
the conventional studies try to reduce the number of the channel parameters  by the following three
approaches: (1)  Gauss-Markov model \cite{gao_relay14}, which captures
channel variation through symbol-by-symbol updating; (2) basis expansion model (BEM) \cite{GiannakisBEM}, which
decomposes channels into the superposition of time-varying
basis functions weighted by time-invariant coefficients; (3) known temporal channel covariance matrix, whose most
dominant eigenvectors can act as basis vectors  to span the time varying channels. Among these approaches, BEM approach
 has attracted most attentions due to its easier implementation, while channel
 covariance matrix approach suffers from huge complexity and overhead cost.

Similarly, for massive MIMO system with closely equipped array antennas, the channels are also highly correlated
in the spatial domain. Based on this fact, \cite{Caire} and \cite{yin} assume the spatial channel covariance matrix is known and
use the dominant eigenvectors to span the spatial channels. A natural question then arises: does there exist a simpler counterpart of
temporal BEM to reduce the spatial channel dimension?

Motivated by this, we propose a discrete Fourier transform (DFT) based spatial basis expansion model (SBEM)
for massive uniform linear array (ULA). Meanwhile, we also jointly consider the temporal basis expansion model under
time selective environment, resulting into the  spatial-temporal BEM (ST-BEM) that could efficiently reduce the channel
parameters both in time and spatial domains. Importantly, the proposed framework is suitable for both TDD and FDD systems
by exploiting the angle reciprocity, and can be efficiently implemented by
the fast Fourier transform (FFT) and partial FFT. Various numerical results are provided to corroborate the proposed scheme.

\enlargethispage{-0.3cm}
\section{System Model and Channel Characteristics}\label{sec:modelandproperty}
\subsection{System Model}\label{sec:system model}

\begin{figure}[t]
      \centering
     \includegraphics[width=85mm]{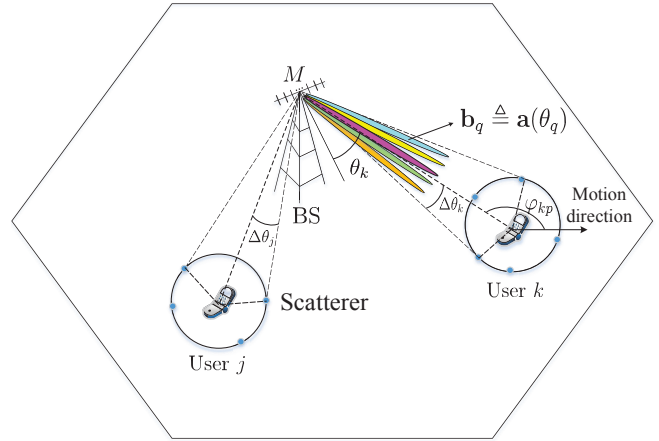}
     \caption{System model. Users are surrounded by $P$ local scatterers and
     the mean DOA and AS of user-$k$ are $\theta_k$ and $\Delta\theta_k$, respectively. When users move around
     in the circle, the spatial AS seen by BS is generally unchanged.}
     \label{fig:systemmodel}
     \vspace{-0.5em}
\end{figure}
Let us consider a multiuser massive MIMO system, where BS is equipped with $M\!\gg\! 1$
antennas in the form of ULA serving $K$ single-antenna users and each user is
surrounded by a circle of $P$ local scatterers,  as shown in Fig. \ref{fig:systemmodel}.
Considering the motion of users, the baseband channels between users and BS
are assumed
to be time-selective flat-fading.
Note that as users move around within the circular area,
the spatial directions of users seen by the BS can be viewed as unchanged, until
users' locations have significantly changed.
Hence, the propagation from user-$k$ to BS is assumed to consist of $P\gg1$ rays and
the corresponding $M\times 1$ uplink channel at time index $n$ can be expressed as \cite{yin,chengshanxiao}:
\begin{equation}\label{equ:channelmodel}
\!\mathbf h_{k}(n)=\frac{1}{\sqrt{P}}\sum_{p=1}^P \alpha_{kp}e^{-j(2\pi f_d nT_s\cos\varphi_{kp}+\phi_{kp})}\mathbf a(\theta_{kp}),
\end{equation}
for $n=0,\ldots, N{-}1$, where $\alpha_{kp}$ denotes the time-varying complex gain
of the $p$-th ray; $f_d$ is the maximum Doppler frequency;
$T_s$ is the system sampling period;
$\varphi_{kp}$ is the angle between the  user-$k$'s uplink transmitted signal and
its motion direction (see Fig. 1);
$\phi_{kp}$ signifies the initial phase, which is uniformly distributed in $[0,2\pi]$;
$\alpha_{kp}$'s, $\varphi_{kp}$'s and $\phi_{kp}$'s are independent and identical distributed (i.i.d.) among different rays.
Moreover, $\mathbf a(\theta_{kp})\in\mathbb{C}^{M\times 1}$ is the array manifold vector defined as
\begin{equation}\label{equ:steeringvector}
\mathbf a(\theta_{kp})=\left[1,e^{j\frac{2\pi d}{\lambda}\sin\theta_{kp}},\ldots,e^{j\frac{2\pi d}{\lambda}(M-1)\sin\theta_{kp}}\right]^T,
\end{equation}
where $d$ is the antenna spacing, $\lambda$ is the signal wavelength, and $\theta_{kp}$ is
the direction of arrival (DOA) of the $p$-th ray seen by the BS array.

Similar to \cite{Caire} and \cite{yin}, the incident angular spread (AS) of user-$k$  with mean DOA
$\theta_k$ seen by BS is assumed to be limited in a narrow region, i.e., $[\theta_k{-}\Delta\theta_k,\theta_k{+}\Delta\theta_k]$.
And this spatial AS of each user is generally unaltered when the user moves within the circular area, see Fig. \ref{fig:systemmodel}.
Hence, there exists high correlations among $\mathbf{a}(\theta_{kp})$, $p=1,{\ldots}, P$,
and $\mathbf{h}_k(n)$ can  be expanded from some orthogonal basis as
\begin{align}\label{equ:SBEM}
\mathbf{h}_k(n)=\sum_{q=1}^\tau \psi_{k,q}(n) \mathbf{b}_q, \ 0\leq n\leq N-1.
\end{align}
As long as we find a set of uniform basis vectors $\mathbf{b}_q$'s
(i.e., the colored beams in Fig. \ref{fig:systemmodel}) for any possible $\mathbf{h}_k(n)$ and
for any time $ 0\leq n\leq N-1$,
then the task of channel estimation will be greatly simplified to estimating $\tau~(\ll M)$ expansion coefficients $\psi_{k,q}(n)$'s only.

To capture the rapid variation of $\psi_{k,q}(n)$'s,
the complex componential basis expansion model (CE-BEM) \cite{GiannakisBEM} is applied in this paper so
that during any time interval of $NT_s$, $\psi_{k,q}(n)$'s can be
modeled as \cite{gao_relay,wenxian(1)}
\begin{align}\label{equ:TDBEM}
  \psi_{k,q}(n)=\sum_{r=0}^R\lambda_{k,q}^{r}e^{j2\pi(r-R/2)n/N}, \ \ 0\leq n\leq N-1,
\end{align}
where $\lambda_{k,q}^r$'s are the CE-BEM coefficients
that remain invariant within one interval of $NT_s$ but may
vary in the next interval. Note that the order of the bases $R$ is a function of the
channel bandwidth and the interval length.

\subsection{Characteristics of Channels in Space and Time Domain}\label{sec:ULAcharacteristics}
Define the normalized DFT of the channel vector $\mathbf{h}_k(n)$
as $\tilde{\mathbf h}_k(n)=\mathbf{F}\mathbf{h}_k(n)$, where $\mathbf F$ is the $M\times M$ DFT matrix
whose $(p,q)$th element is  $\left[\mathbf F\right]_{pq}=e^{-j\frac{2\pi}{M}pq}/\sqrt{M}$.

\begin{property} \label{property:2}
For the channel model \eqref{equ:channelmodel},  $\tilde{\mathbf h}_k(n)$
is approximately a sparse vector with most channel power concentrated on few DFT entries.
\end{property}
\begin{proof}
Based on the Vandermonde structure of $\mathbf a(\theta_{kp})$, the property is comprehensible. To illustrate this,
let us define $\mathcal{B}_k$ as the index set of the continuous DFT points that contain at least $\eta$ portion of the total channel power.
Then the left bound of $\mathcal{B}_k$ is determined by the DFT of the leftmost ray with $\theta_{kp}=\theta_k-\Delta\theta_k$ and
can be expressed as $\lfloor M\frac{d}{\lambda}\sin(\theta_k-\Delta\theta_k)\rfloor-\lceil B_{\max}/2\rceil $, where
$B_{\max}$ is the upper bound of the cardinality of $\mathcal{B}_k$, i.e., $|\mathcal{B}_k|$, corresponding to any single-ray case with DOA inside $[\theta_k-\Delta\theta_k,\theta_k+\Delta\theta_k]$,
and $\lceil \cdot\rceil$ and $\lfloor \cdot\rfloor$ denote the integer ceiling and integer floor, respectively.
Similarly, the right bound of $\mathcal{B}_k$ depends on the DFT of the rightmost ray with $\theta_{kp}=\theta_k+\Delta\theta_k$ and can be expressed as
$\lceil M\frac{d}{\lambda}\sin(\theta_k+\Delta\theta_k)\rceil+\lceil B_{\max}/2 \rceil$.
Then for any single ray with incident DOA inside $[\theta_k-\Delta\theta_k,\theta_k+\Delta\theta_k]$,
$|\mathcal{B}_k|$ can be approximated as
\begin{align}\label{equ:multi-rayrange}
       |\mathcal{B}_k|&{\approx} \lceil M\frac{d}{\lambda}\sin(\theta_k{+}\Delta\theta_k) \rceil
      {-}\lfloor M\frac{d}{\lambda}\sin(\theta_k{-}\Delta\theta_k) \rfloor {+}1{+}B_{\max}\notag \\
       &{\approx} \lceil2M\frac{d}{\lambda}\cdot|\cos\theta_k|\cdot\Delta\theta_k+1\rceil+B_{\max}.
\end{align}
To demonstrate this, an example of a $9$-ray channel with AS $[25^\circ,29^\circ]$ is given in Fig. \ref{fig:multi-raypower},
where the discrete time Fourier transform (DTFT) of each single ray as well as the DFT of overall multi-rays
are depicted respectively. This numerical simulation shows that  the actual cardinality of $\mathcal{B}_k$
containing $\eta=95\%$ power is $15$, which is just equal to the result of \eqref{equ:multi-rayrange}.

\begin{figure}
\centering
\includegraphics[width=85mm,height=50mm]{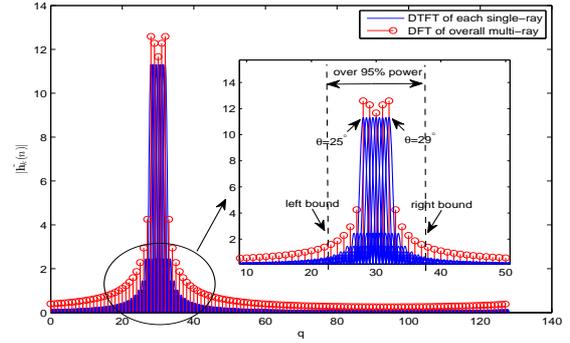}
\vspace{-1em}
\caption{Example of the DFT of channel model $\mathbf h_k(n)$ with DOA inside $[25^\circ,29^\circ]$ and $M=128, d=\lambda/2$.
 Since the AS is unchanged for the whole interval, the position of $\mathcal{B}_k$ is also the same for
 all $0\leq n\leq N-1$. \label{fig:multi-raypower}}
\end{figure}

Based on \eqref{equ:multi-rayrange} and recalling that $\Delta\theta_k$ is small, it is obvious  that
$|\mathcal{B}_k|{\approx} \left\lceil\frac{2 Md}{\lambda}\cdot|\cos\theta_k|\cdot\Delta\theta_k+1\right\rceil{+} B_{\max}$
is still small compared to $M$. Hence, $\tilde{\mathbf h}_k(n)$ is approximately sparse with most power being  contained in limited number of entries.
\end{proof}

As per \emph{Property \ref{property:2}}, the key idea of this paper is to approximate the channel vector with fewer parameters as
\begin{align}\label{equ:preambleapprox}
\!\!\mathbf h_{k}(n)&{=}\mathbf F^H\tilde{\mathbf h}_{k}(n)
  {\approx}\!\left[\mathbf F^H\right]_{:,\mathcal{B}_{k}}\!\!\left[\tilde{\mathbf h}_{k}(n)\right]_{\mathcal{B}_{k},:}
    \!{=}\!\!\sum_{q\in\mathcal{B}_k}\!\!\tilde{h}_{k,q}(n)\mathbf f_q,
\end{align}
where $\tilde{h}_{k,q}(n)\triangleq [\tilde{\mathbf h}_k(n)]_q$ denotes the $q$-th element of $\tilde{\mathbf h}_k(n)$ while $\mathbf f_q$
is the $q$-th column of $\mathbf F^H$; $[\cdot]_{:,\mathcal{B}_k}$ and $[\cdot]_{\mathcal{B}_k,:}$ denote the
sub-matrices by collecting columns or rows indexed by $\mathcal{B}_k$, respectively. Compare with \eqref{equ:SBEM},
the expansion in \eqref{equ:preambleapprox} is in the form of BEM where the basis vectors, $\mathbf{b}_q\triangleq\mathbf{f}_q$, are orthogonal to each other. Hence,  we only need to estimate the limited BEM coefficients $\tilde{h}_{k,q}(n)$.

Interestingly, the DFT vector $\mathbf f_q$ coincides with the steering vector as
$\mathbf f_q=\mathbf a(\theta_q)$ where $\theta_q=\arcsin\frac{q\lambda}{Md}$, which means that
$\mathbf f_q$ formulates an array beam towards
the physical direction $\theta_q=\arcsin\frac{q\lambda}{Md}$ (see Fig. \ref{fig:systemmodel}). Hence, all
 beams $\mathbf f_q$'s inside $\mathcal{B}_k$ will point towards the AS of user-$k$ and are orthogonal to each other. Consequently,
the beam indices, i.e., $\mathcal{B}_k$, are defined as the \emph{spatial signature} of user-$k$, and \eqref{equ:preambleapprox} can be deemed as the
\emph{spatial BEM} (SBEM).

Note that, since the spatial AS for users are unchanged for $0\leq n\leq N{-}1$, their spatial signature
$\mathcal{B}_k$ will also be the same for the whole interval. This is illustrated in
Fig. \ref{fig:multi-raypower}, where the position of $\mathcal{B}_k$ of $\tilde{\mathbf h}_k(n)$ is almost the same
for all time index $ 0\leq n\leq N{-}1$.

%Since that practical standards normally regulate a fixed number of channel parameters instead of a dynamic number $|\mathcal{B}_k|$,
%let us denote $\tau$ as the number of the channel parameters that systems could handle, and
%then we should select $\mathcal{B}_k$ such that $ [\tilde{\mathbf h}_k(n)]_{\mathcal{B}_k,:}$ possesses the maximum channel power, i.e.,
%\begin{align}\label{equ:phaserotationobjective}
%  \max_{\mathcal{B}_k}\ \ &\left\|[\tilde{\mathbf h}_k(n)]_{\mathcal{B}_k,:}\right\|^2
%  \ \ \ \ \tu{subject to} \  |\mathcal{B}_k|=\tau.
%\end{align}
%The above optimization can be achieved simply by sliding window of size $\tau$
%over the elements in $\tilde{\mathbf h}_k(n)$.

\begin{property}\label{property:bandlimited}
  All the components of $\tilde{\mathbf h}_k(n)$, say $\tilde{h}_{k,q}(n), q\!=\!0,{\ldots},M{-}1$, are
  band-limited and the maximum bandwidth of the power spectra is exactly equal to $f_d$.
\end{property}
\begin{proof}
  First let us define the time-domain discrete correlation matrix of $\mathbf h_k(n)$ as $\mathbf R_k(m)\triangleq \mathbb E\{\mathbf h_k(n)\mathbf h_k^H(n+m)\}$
  and then the $(i,l)$th components of $\mathbf R_k(m)$ is given as
  \begin{align} \label{equ:correlation}
    [\mathbf R_k(m)]_{i,l}&=\mathbb E\{h_{k,i}(n)h^*_{k,l}(n+m)\}\notag\\
    &=\frac{1}{P}\sum_{p=1}^P \mathbb E\{|\alpha_{kp}|^2\}\mathbb E\{e^{-j2\pi f_d mT_s \cos\varphi_{k,p}}\}\notag\\
    & \ \ \cdot \mathbb E\{e^{j\frac{2\pi d}{\lambda}(i-l)\sin\theta_{kp}}\},
  \end{align}
  where $\mathbb E\{\cdot\}$ denotes expectation.
%\textcolor[rgb]{1.00,0.00,0.00}{  Considering the rich local scatters surrounding the users and the high random mobility of users,
%  {强调不同，基站端不变，用户端丰富，跟传统不一样的，强调不管什么R都给出一个带限}}

  It is worth noticing that different from the conventional Clarke's reference model \cite{chengshanxiao}, where
  the incident angles of users' signals seen at the BS are assumed to be uniformly distributed on $[0,2\pi]$,
  the spatial AS of user-$k$ here is unchanged within a narrow range, i.e., $[\theta_k-\Delta\theta_k,\theta_k+\Delta\theta_k]$
  as discussed before, while the angle $\varphi_{kp}$ at the user end is assumed to be randomly distributed
  over $[0,2\pi]$ for the rich local scatters surrounding the users and the high random mobility of users.
  Therefore, the expectation in \eqref{equ:correlation} is mainly focused on $\varphi_{kp}$ and then we have
  \begin{align}\label{equ:timecorrelation}
    [\mathbf R_k(m)]_{i,l}&=J_0(2\pi f_d m T_s)\cdot g(2\pi d/\lambda(l-i)),
  \end{align}
  where $J_0(x)=\frac{1}{2\pi}\int_{-\pi}^\pi e^{-jx\cos y}dy$ is the zero-order Bessel function of the first kind,
  and $g(\cdot)$ is similarly defined as
  $g(x)=\frac{1}{2\Delta\theta_k}\int_{\theta_k-\Delta\theta_k}^{\theta_k+\Delta\theta_k}e^{-jx\sin y}dy$.
  From \cite{Clarke}, the power spectrum of $J_0(2\pi f_d m Ts)$ is the well-known ``U-shape'' function, namely,
  $S_{J_0}(f)=\frac{1}{\pi f_d\sqrt{1-f^2/f_d^2}}, f\in[-f_d,f_d]$,
  Therefore, the bandwidth of \eqref{equ:timecorrelation} is upper bounded by $f_d$.
  Moveover, it also shows that no matter what the range of the spatial AS is, it will not affect
  this time-domain bandwidth of $\tilde{h}_{k,q}(n)$.

  Since that $\tilde{h}_{k,q}(n)=\sum_{i=0}^{M-1}h_{k,i}(n)e^{-j\frac{2\pi}{M}iq}$, we then have
  \begin{align}
    &\mathbb E\{\tilde{h}_{k,q}(n)\tilde{h}_{k,q}^*(n+m)\}\notag\\
=&\sum_{i=0}^{M-1}\sum_{l=0}^{M-1}\mathbb E\{h_{k,i}(n)h_{k,l}^*(n+m)\}e^{-j\frac{2\pi }{M}q(i-l)}.
  \end{align}
This correlation function is the superimposition of multiple band-limited signals and thus the property
is obvious.
\end{proof}

\begin{figure}
\centering
\includegraphics[width=85mm]{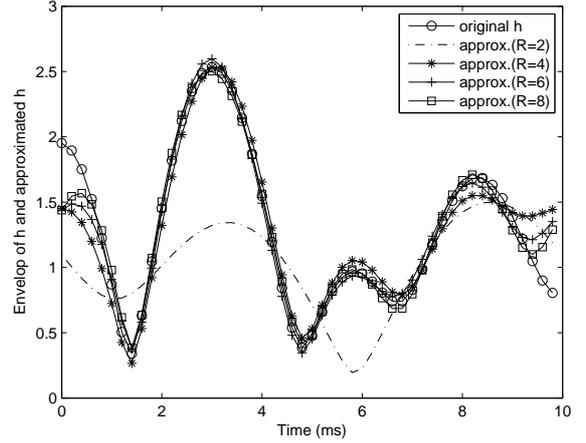}
\vspace{-1em}
\caption{CE-BEM approximation of $\tilde{h}_{k,2}(n), n=1,\ldots,N$ with different values of $R$, where
$M=128,d=\lambda/2, f_d=200$Hz, $T_s=0.1$ms and $N=100$.
\label{fig:BEMapproximate}}
\end{figure}

Following \emph{Property \ref{property:bandlimited}}, the CE-BEM approximation of band-limited signals in \eqref{equ:TDBEM}
is then justified and the order $R$ should be at least  $2\lceil f_dNT_s\rceil$ in order to provide sufficient degrees of
freedom \cite{gao_relay}. To illustrate this, an  approximation example of $\tilde{h}_{k,2}(n)$ is given in Fig. \ref{fig:BEMapproximate}, where
the simulation parameters are taken as $M=128,d=\lambda/2, f_d=200$Hz, $T_s=0.1$ms and $N=100$. It can be seen that
when $R\geq 2\lceil f_dNT_s\rceil=4$, the CE-BEM approximation of $\tilde{h}_{k,2}(n)$ is pretty good.
However, for $R=2$ the ambiguous estimation appears due to the lack of sufficient sampling degrees of freedom.

Combining above two properties, the approximation of massive MIMO time-varying channels can be simplified as
\begin{align}\label{equ:STBEM}
  \mathbf h_k(n) &\approx \sum_{q\in\mathcal{B}_k} \tilde{h}_{k,q}(n)\mathbf f_q=
  \sum_{q\in\mathcal{B}_k} \sum_{r=0}^R \lambda_{k,q}^r e^{j2\pi(r-R/2)n/N}\mathbf f_q\notag\\
  &=\sum_{q\in\mathcal{B}_k} \bm{\lambda}_{k,q}^T \mathbf c_n \mathbf f_q, \ \ \  n=0,\ldots,N-1,
\end{align}
where $\bm \lambda_{k,q}{=}[\lambda_q^0,\ldots,\lambda_q^R]^T$ and
$\mathbf c_n{=}[e^{-j\frac{2\pi n}{N}\frac{R}{2}},\ldots,e^{j\frac{2\pi n}{N}\frac{R}{2}}]^T$.
Generally, \eqref{equ:STBEM} can be viewed as a joint \emph{spatial-temporal BEM} (ST-BEM) for
massive MIMO fast-fading channels.

%\begin{remark}
%Although high mobility of users will induce fast-fading channel gains, DOA and AS of a user can still be treated
%as unchanged within several or even tens of the channel sampling intervals for users may not
%physically change their positions in such a short time at the level of millisecond.
%Hence, the spatial signatures  $\mathcal{B}_k$
%of each user could be deemed as unaltered such that \eqref{equ:STBEM}
%holds true for $n=0,\ldots,N-1$.
%\end{remark}

\section{Channel Estimation Scheme With ST-BEM }\label{sec:Uplinktraining}

In this section, we consider the uplink/downlink transmission
that utilizes the spatial signatures to realize the orthogonal training
among  different users with much reduced overhead. The transmissions
between BS and users always start from an uplink preamble to obtain
the spatial signature of each user. Then users are grouped for
uplink/downlink training based on their spatial signatures.

The pilot symbol aided modulation (PSAM) technique \cite{PSAM}
is employed to probe the uplink/downlink time-varying channels, where pilot symbols are inserted
among information symbols in each interval of $NT_s$. Let us define $\mathcal{T}_t=\{n_0,n_1,\ldots,n_{T-1}\}\subset\{0,\ldots,N-1\}$
as the time index set for pilot symbols.

\subsection{Obtain Spatial Information through Uplink Preamble}

Since the AS information for all users will keep unchanged for a relative long time,
then users can be scheduled during the preamble period to
yield the initial channel estimate $\hat{\mathbf h}_k^{\textup{pre}}$ for $k=1,{\ldots},K$, respectively,
by using existing conventional uplink channel estimation methods, e.g. least square (LS) and minimum mean square error (MMSE).
And the next step is  to extract the spatial signature $\mathcal{B}_k$ of size $\tau$
that contains the maximum power of $\mathbf F\hat{\mathbf h}_k^{\textup{pre}}$ for each user.

\subsection{Uplink Training with User Grouping}
Keep in mind that the non-overlapped properties
of different users' spatial signatures could be utilized to release
the pressure of training overheads.
Let us divide users into separate groups according to their spatial signatures.
Specifically, users are allocated to the same group if their spatial signatures do not
overlap, i.e., $\mathcal{B}_k\cap\mathcal{B}_l=\emptyset$.
Assume that all users are divided into $G$ groups
and denote the user index set of the $g$-th group as $\mathcal{U}_g$.
We expect that $G\ll K$ since the cardinality of $\mathcal{B}_k$, namely, the number
of selected components of channels, is set as $\tau=|\mathcal{B}_k|$ which is
much smaller than $M$. Meanwhile, users are randomly distributed in the service region such that
their spatial signatures will also be randomly distributed.

As channels of different users in the same group could be discriminated by their spatial
signatures, we could assign the same pilot sequence to all users in one group.
Then let us specify the received training signals at BS as
$\mathbf Y=[\mathbf y(n_0),\mathbf y(n_1),\ldots, \mathbf y(n_{T-1})]\in\mathbb{C}^{M\times T}$
and the common transmitted pilot symbols of users in $g$-th group
as $\mathbf S_g=\textup{diag}\{s_g(n_0),s_g(n_1),\ldots,s_g(n_{T-1})\}$ with $\sum_{i=0}^{T-1}|s_g(n_i)|^2=1$. Then we have
\begin{align}\label{equ:ULtraining}
\mathbf Y&=\sum_{g=1}^G\sum_{k\in\mathcal{U}_g}
 \left[\mathbf h_k(n_0), {\ldots},\mathbf h_k(n_{T\!-\!1})\right]\sqrt{P_{k}^{\textup{ul}}}\mathbf S_{g}+\mathbf N\notag\\
  &=\sum_{g=1}^G\sum_{k\in\mathcal{U}_g} \sqrt{P_{k}^{\textup{ul}}}\mathbf F^H\bm \Lambda_k \mathbf C \mathbf S_g +\mathbf N\notag\\
  &=\mathbf F^H\left[\sum_{k\in\mathcal{U}_1}\!\!\sqrt{P_{k}^{\textup{ul}}}\bm \Lambda_k,{\ldots},\sum_{k\in\mathcal{U}_G}\!\!\sqrt{P_{k}^{\textup{ul}}}\bm \Lambda_k\right]\notag\\
 &\ \cdot \left[(\mathbf C\mathbf S_1)^H,{\ldots}, (\mathbf C\mathbf S_G)^H\right]^H+\mathbf N\notag\\
  &=\mathbf F^H\bm \Lambda \left[(\mathbf C\mathbf S_1)^H,{\ldots}, (\mathbf C\mathbf S_G)^H\right]^H+\mathbf N
\end{align}
where $P_{k}^{\textup{ul}}$ is the uplink power constraint at user-$k$;
$\bm \Lambda_k=[\bm\lambda_{k,0},\bm \lambda_{k,1},\ldots,\bm \lambda_{k,M-1}]^T$
denotes the CE-BEM coefficients for user-$k$; %$\bm \Lambda=[\sum_{k\in\mathcal{U}_1}\!\!\sqrt{P_{k}^{\textup{ut}}}\bm \Lambda_k,{\ldots},\sum_{k\in\mathcal{U}_G}\!\!\sqrt{P_{k}^{\textup{ut}}}\bm \Lambda_k]$;
$\mathbf C=[\mathbf c_{n_0},\mathbf c_{n_1},\ldots,\mathbf c_{n_{T-1}}]$
 and $\mathbf N$ is the noise matrix whose elements are  i.i.d. $\mathcal{CN}(0,\sigma_n^2)$.

When $T\geq G(R+1)$, there will be adequate observations to estimate all the unknowns in $\bm \Lambda$.
In this paper, we resort to the LS estimator as \cite{gao_relay} did and then obtain
\begin{align}
  \hat{\bm \Lambda} = \mathbf F\mathbf Y\left\{\left[(\mathbf C\mathbf S_1)^H,{\ldots}, (\mathbf C\mathbf S_G)^H\right]^{H}\right\}^{\dag},
\end{align}
and the mean square error (MSE) is given as
\begin{align}\label{equ:ULMSE}
  \mathbb E\{\|\hat{\bm \Lambda}-\bm \Lambda\|_F^2\}=&M\sigma_n^2\ \textup{tr}\big\{\big(\left[(\mathbf C\mathbf S_1)^H,{\ldots}, (\mathbf C\mathbf S_K)^H\right]^H\notag\\
  &\left. \cdot\left[(\mathbf C\mathbf S_1),{\ldots}, (\mathbf C\mathbf S_K)\right]\big)^{-1}\right\}
\end{align}
In line with the idea of \cite{gao_relay}, we know that to minimize the MSE in \eqref{equ:ULMSE},
the optimal pilot symbols adopted by users should satisfy the following constraints as
\begin{align}\label{equ:ULoptimaldesign}
  \mathbf C\mathbf S_g\mathbf S_g^H\mathbf C^H=\mathbf I_{R+1},\
  \mathbf C\mathbf S_g\mathbf S_{g'}^H\mathbf C^H=\mathbf 0_{R+1}, \ \forall\ g\neq g'.
\end{align}
Moreover,
the optimal pilot symbols for different groups are proved to be equi-powered, equi-spaced over $\{0,\ldots,N-1\}$,
and phase shift orthogonal. One example of such kind of
pilot sequences is
\begin{align}\label{equ:optimalsequence}
  s_g(n_i)=&\sqrt{1/T}e^{j2\pi i (g-1)(R+1)/T},\notag\\
  & g=1,\ldots,G; i=0,\ldots,T-1.
\end{align}
Let us then focus on user-$k$ in group-$g$ and get
\begin{align}
  \hat{\bm \Lambda}_k = \bm \Lambda_k+\sum_{l\in\{\mathcal{U}_g\backslash k\}}\sqrt{P_{l}^{\textup{ul}}/P_{k}^{\textup{ul}}}\bm \Lambda_l+
  \frac{1}{\sqrt{P_{k}^{\textup{ul}}/\sigma_n^2}}\mathbf N_k,
\end{align}
where $\mathbf N_k{\in}\mathbb{C}^{M\times R+1}$ has the i.i.d. $\mathcal{CN}(0,1)$ elements.
Considering the disjoint spatial signatures of users in the same group, we can straightforwardly extract
\begin{align}\label{equ:ULlambda_hat}
   \widehat{\left[\bm \Lambda_k\right]}_{\mathcal{B}_k,:}&=[\hat{\bm \Lambda}_k ]_{\mathcal{B}_k,:} =\left[\bm \Lambda_k\right]_{\mathcal{B}_k,:}
+\!\!\sum_{l\in\{\mathcal{U}_g\backslash k\}}\sqrt{P_{l}^{\textup{ul}}/P_{k}^{\textup{ul}}}\left[\bm \Lambda_l\right]_{\mathcal{B}_k,:}\notag\\
&+1/\sqrt{P_{k}^{\textup{ul}}/\sigma_n^2}\left[\mathbf N_k\right]_{\mathcal{B}_k,:},
\ \forall\  k\in\mathcal{U}_g.
\end{align}
Bearing in mind that $\mathcal{B}_l$ and $\mathcal{B}_k$ are kept away from each other,
we know the entries of $\left[\bm \Lambda_l\right]_{\mathcal{B}_k,:}$
in \eqref{equ:ULlambda_hat} are negligible such that the pilot contamination term caused by reusing the same pilot
in one group is immediately reduced, and then $\bm \Lambda_k$ for user-$k$ can be approximated as
\begin{align}
  \bm \Lambda_k=\left[\mathbf 0^T\ \widehat{\left[\bm \Lambda_k\right]}_{\mathcal{B}_k,:}^H\ \mathbf 0^T\right]^H,
\end{align}
where the two all-zero matrices $\mathbf 0$ have appropriate sizes. With this estimated coefficients in hand,
the channel estimate $\hat{\mathbf h}_k(n)$ thus can be obtained by \eqref{equ:STBEM}.
\begin{remark}
  By grouping users according to their spatial signatures, the total pilot overheads can be reduced significantly
  from $T=K(R+1)$ to $T=G(R+1)$ with $G\ll K$. Furthermore, the operations in \eqref{equ:preambleapprox} and \eqref{equ:STBEM} can
  be accelerated by partial fast Fourier transform (FFT) \cite{partialFFT1}, which further moderates the high calculation complexity.
\end{remark}

\subsection{Downlink Channel Representation with Angle Reciprocity}\label{sec:downlinktraining}
%DL training is necessary for time-selective channels.
 Denote the downlink channel from BS to  user-$k$ as $\mathbf g_k^H(n)\in\mathbb{C}^{1\times M}$.
 Similar to  \eqref{equ:STBEM},
$\mathbf g_k(n)\in\mathbb{C}^{M\times 1}$ can be modeled as
 \begin{align}\label{equ:DLSTBEM}
   \mathbf g_k(n)\approx \sum_{q\in\mathcal{B}'_k} \tilde{g}_{k,q}(n)\mathbf f_q= \sum_{q\in\mathcal{B}'_k} \bm{\lambda}_{k,q}^{d^T} \mathbf c_n \mathbf f_q,
 \end{align}
where $\bm \lambda_{k,q}^{d}$ denotes the CE-BEM coefficients of downlink channels and
$\mathcal{B}'_k$ is the downlink version of spatial signatures.
All other parameters have been defined in \eqref{equ:ULtraining}.
Similar to the uplink, once $\mathcal{B}'_k$ is determined, the downlink channel estimation for $\mathbf g_k(n)$ will be
simplified to estimating those remaining unknown coefficients $\bm \lambda_{k,q}^{d}$'s.

To determine $\mathcal{B}'_k$, let us first introduce an important property of wireless channels.
Since the propagation path of electromagnetic wave is reciprocal, we know that only the signal
wave that physically reverses the uplink path can reach the user during the downlink period.
Hence, downlink signals that could effectively arrive at the user should have the same DOD spread as the uplink DOA spread.
We call this property as  the \emph{angle reciprocity}.
Similar assumptions have already been directly adopted in many existing works, such as \cite{DOAreciprocity1 ,DOAreciprocity3}.

Based on the angle reciprocity and bearing in mind that
the spatial signatures are exactly determined by the AS, we know that
$\mathcal{B}'_k$ can be directly determined by $\mathcal{B}_k$.
Specifically, according to \emph{Property \ref{property:2}}, we will have
\begin{align}
\sin\theta_{k} = \frac{q\lambda_1}{Md}=\frac{q'\lambda_2}{Md},\ \textup{with}\ q\in\mathcal{B}_k,\
q'\in\mathcal{B}'_k,
\end{align}
where $\lambda_1$ and $\lambda_2$ denote the uplink/downlink carrier wavelengths, respectively.
Then the integer set $\mathcal{B}'_k$ can
be expressed as $\mathcal{B}'_k=\{q'_{\min},q'_{\min}+1,\ldots,q'_{\max}\}$ with
\begin{align}\label{equ:DLBk}
  q'_{\min}=\left\lfloor\frac{\lambda_1}{\lambda_2}q_{\min}\right\rfloor,\ \ q'_{\max}=\left\lceil\frac{\lambda_1}{\lambda_2}q_{\max}\right\rceil,
\end{align}
where $q_{\min}\leq q\leq q_{\max}, \forall~ q\in\mathcal{B}_k$.

%Therefore,  $\mathbf g_k^H{\in}\mathbb{C}^{1\times M}$ can be approximated  by SBEM with the same
%spatial signatures $\mathcal{B}_k$ of $\mathbf{h}_k(n)$, i.e.,
%\begin{align}\label{equ:downlinkbeamchannel}
%  \mathbf g_{k}&=\mathbf F^H\tilde{\mathbf g}_{k}
%\approx\left[\mathbf F^H\right]_{:,\mathcal{B}_k}\left[\tilde{\mathbf g}_{k}\right]_{\mathcal{B}_k,:}=\sum_{q\in\mathcal{B}_k}\tilde{g}_{k,q}\mathbf f_q,
%\end{align}
%where $\tilde{g}_{k,q}\triangleq[\tilde{\mathbf g}_k]_q$ denotes the $q$-th element
%of $\tilde{\mathbf g}_k=\mathbf F\mathbf g_k$.
%Following \eqref{equ:downlinkbeamchannel}, the estimation of DL channels $\mathbf{g}_k$ is simplified to estimate only
%$\tau$ unknown SBEM coefficients $\tilde{g}_{k,q}$.

%\begin{remark}
% Note that the angle reciprocity of UL DOA and DL DOD is straightforwardly true for TDD systems.
% For FDD systems, if  the frequency of the DL channel is  not too far from that of
% the UL channel, e.g. less than several GHz,  then the reciprocity
% between DOD and DOA still accurately holds.
% The reason is that for the typical transmission environment, the relative permittivity and the conductivity of the obstacles do not change
%in the scale of  several dozens of GHz \cite{EMproperty}.
%Therefore,  the proposed training strategy is suitable for both TDD and FDD massive MIMO systems in time-varying environments.
%\end{remark}

\subsection{Downlink Training with User Grouping}\label{sec:downlinkgrouping}
Following \eqref{equ:DLSTBEM}, the effective dimensions of downlink channels for all users have been reduced to $\tau\ll M$,
and thus by adopting the uplink user grouping strategy directly, users in each groups can be simultaneously
scheduled for downlink transmission with only $T=\tau(R+1)$ pilot symbols required.

Similar to the uplink training procedures, we have the received $\mathbf y^d_k=[y_k(n_0),\ldots,y_k(n_{T-1})]^T$ at user-$k$ in $\mathcal{U}_g$ as
\begin{align}\label{equ:DLtraining}
  \mathbf y^d_k=&\sum_{l\in\mathcal{U}_g}\sum_{1\leq i\leq \tau, q_i\in\mathcal{B}_l}\sqrt{P_l^{\textup{dl}}/\tau}\mathbf S^d_i\mathbf C^T\bm\lambda_{k,q_i}^d+\mathbf n_k\notag\\
  =&\sum_{l\in\mathcal{U}_g}\left[\mathbf S^d_1\mathbf C^T,\ldots,\mathbf S^d_{\tau}\mathbf C^T\right]
  \textup{vec}\left(\left[\bm\Lambda_k^{d}\right]_{\mathcal{B}_l,:}^T\right)+\mathbf n_k,
\end{align}
where diagonal matrices $\mathbf S^d_i\in\mathbb{C}^{T\times T},i=1,\ldots,\tau$ denote the transmitted pilot sequence for all users in $\mathcal{U}_g$
at their own $\tau$ beam directions $\mathbf f_q$'s respectively, and $P_l^{\textup{dl}}$ is the total downlink power constraint for user-$l$.
Moreover, $\bm\Lambda_k^{d}$ is the downlink version of $\bm\Lambda_k$ and noise vector satisfies
$\mathbf n_k\sim\mathcal{CN}(\mathbf 0,\sigma_n^2\mathbf I_T)$.

Like \eqref{equ:optimalsequence}, the optimal pilot sequences $\mathbf S^d_i\in\mathbb{C}^{T\times T},i=1,\ldots,\tau$
for \eqref{equ:DLtraining} are also equi-powered, equi-spaced and
phase shift orthogonal. Then the downlink channels can be recovered similarly, procedures of which are omitted due to space limitations.
To complete the channel estimation, each user only has to feed back $\tau$ components $\widehat{[\bm \Lambda_k^d]}_{\mathcal{B}_k,:}$
to BS.
\begin{remark}
It can be found that user-$k$ does not
need the knowledge of spatial signature set $\mathcal{B}_k$ to perform the estimation of $\widehat{[\bm \Lambda_k^d]}_{\mathcal{B}_k,:}$.
This removes the necessity of feedback from BS to the user and is thus a key advantage that makes the proposed downlink channel estimation strategy suitable for fast-fading environments.
\end{remark}

\section{Simulations}\label{sec:simulation}
In this section, we demonstrate the effectiveness of the proposed strategy through  numerical examples.
We select $M{=}128$, $d{=}\lambda/2$ and consider $K{=}12$ users gathered into
$4$ disjoint clusters in the coverage area.
Channel vectors are formulated  according to \eqref{equ:channelmodel},
where $P{=}100$, $f_d{=}200$ Hz, $T_s{=}1$ us; $\theta_{kp}$ is uniformly distributed inside
$[\theta_k{-}\Delta\theta_k,\theta_k{+}\Delta\theta_k]$, where two-side AS is supposed be $2\Delta\theta_k{=}4^\circ, 12^\circ, 20^\circ$, respectively.
The value of $\tau$ is assumed to be $\tau{=}16$, which is only $1/8$ of the antenna number. Moreover,
we select $R{=}4$, $N{=}K(R+1){=}60$ for uplink and $N{=}M(R+1){=}640$ for downlink, which satisfies the requirement of
$R\geq2\lceil f_dNT_s\rceil$.
The performance metric of channel estimation is the normalized
MSE, i.e.,
\begin{equation*}
  \tu{MSE} \triangleq \frac{\sum_{k=1}^K\sum_{n=0}^{N-1}\left\|\mathbf h_k(n)-\hat{\mathbf h}_k(n)\right\|^2}{\sum_{k=1}^K\sum_{n=0}^{N-1}\left\|\mathbf h_k(n)\right\|^2}.
\end{equation*}

\begin{figure}[t]
\centering
\includegraphics[width=90mm]{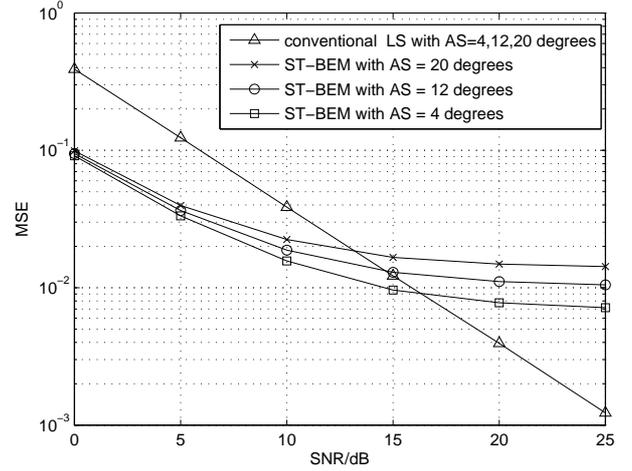}
\vspace{-2em}
\caption{Uplink MSE performance comparison of ST-BEM with $T{=}G(R+1){=}15$ and conventional LS with $T{=}K(R+1){=}60$.
Two-side AS is set as $4^\circ, 12^\circ, 20^\circ$, respectively.
\label{fig:ULMSE_AS}}
\end{figure}

\begin{figure}[t]
\centering
\includegraphics[width=90mm]{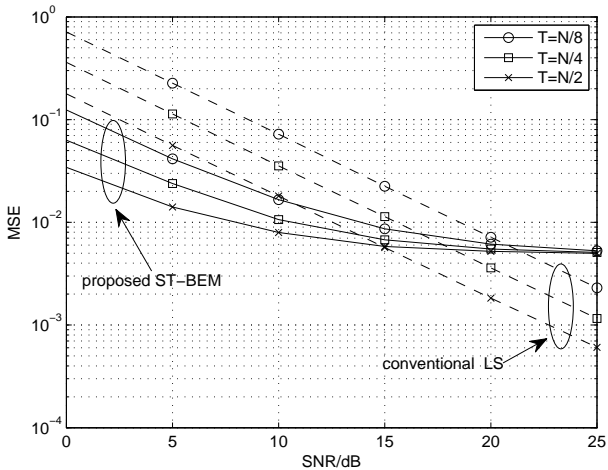}
\vspace{-2em}
\caption{Downlink MSE performance comparison of ST-BEM with $T=N/8, N/4, N/2$, respectively, and conventional LS with $T=M(R+1)=N$.
Two-side AS is set as $4^\circ$.
\label{fig:DLMSE_T}}
\end{figure}

Fig. \ref{fig:ULMSE_AS}
compares the proposed ST-BEM with the conventional LS method for uplink MSE.
To apply the conventional LS, all $N{=}K(R+1){=}60$ symbols are necessary for uplink training,
while only $T{=}G(K+1){=}15$ out of total $N{=}60$ symbols are enough for ST-BEM.
To provide a fair comparison, for any given SNR $\rho$,
the uplink training power for each user is kept the same $P^{\textup{ul}}_k{=}T\rho$  for both methods. Moreover, different values of
AS are also considered.
It can be seen that as the SNR increases, there are error floors for ST-BEM curves  and the error floor is
higher with larger AS.
This phenomenon is not unexpected due to the truncation error of
SBEM from the real channel and  can also be observed in CE-BEM \cite{GiannakisBEM}.
It is seen that when the sufficient length of training is available and when the
computational complexity is acceptable, then the conventional LS method
does not have the error floor for any AS values.
Nevertheless, it can be observed that the channel estimation
from ST-BEM outperforms the conventional method when SNR is relatively low ($\leq \!\! 15$dB). The reasons can be
found from \eqref{equ:ULlambda_hat} where the proposed method only involves $\tau$ components of the noise vector
while the conventional LS method includes the whole noise power.

Fig. \ref{fig:DLMSE_T} compares the downlink MSE performances of ST-BEM with $T{=}N/8, N/4, N/2$, and
conventional LS with all $N=M(R+1)$ symbols. The total power constraint are kept the
same $\sum_{k=1}^KP^{\textup{dl}}_k=KT\rho$ for both methods to ensure the fairness. It can also
be found that the proposed ST-BEM is superior to conventional LS, not only for the higher estimation accuracy in
relative low SNR regions, but also due to the less training overheads, which will help to increase the system
spectral efficiency.

\begin{figure}[t]
\centering
\includegraphics[width=90mm]{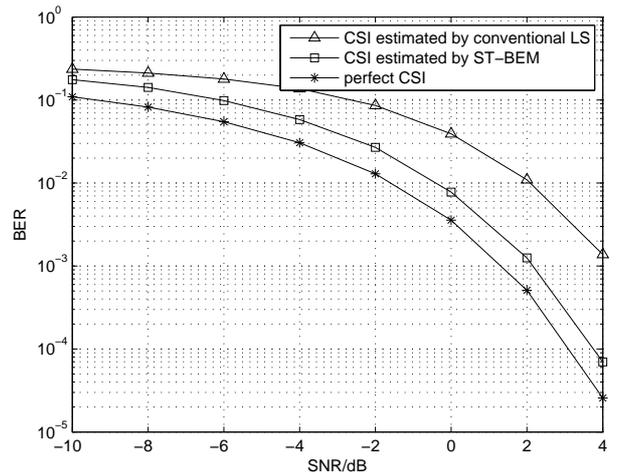}
\vspace{-2em}
\caption{The downlink BER performance comparison of the proposed ST-BEM method with $T=N/8$ and the conventional LS method with $T=N$.
Two-side AS is set as $4^\circ$.
\label{fig:BER}}
\end{figure}

Lastly, we show the bit error rate (BER) performance under QPSK
modulation for the downlink data transmission in Fig. \ref{fig:BER}. Three kinds of CSI are compared,
i.e., perfect CSI, CSI from the proposed ST-BEM, and CSI from the conventional LS.
To keep the comparison fair, the overall training and data transmission power are set as the same for each method.
It is seen that the BER achieved by ST-BEM is better than that of conventional LS and has about 0.5 dB gap from
that of perfect CSI, which corroborates the effectiveness of the proposed ST-BEM.

\section{Conclusions}\label{sec:conclusions}

In this paper, we investigated the uplink/downlink training
for multiuser massive MIMO systems in time-varying environments. We exploited the characteristics
of ULA and proposed a simple DFT-based ST-BEM to represent channel vectors
with reduced parameter dimensions in both spatial and time domains, which helps to reduce the training and feedback overhead significantly.
Meanwhile, the uplink spatial signatures could also be used to simplify the downlink training based on the angle reciprocity,
making the proposed ST-BEM applicable for both TDD and FDD massive MIMO time-varying systems.
Numerical results have demonstrated the effectiveness of the proposed scheme.

\end{document}